\documentclass[aps,pre,twocolumn,showpacs,preprintnumbers,amsmath,amssymb,floatfix]{revtex4-1}
\usepackage{graphicx}
\usepackage{amsfonts}
\usepackage{CJK}

\begin{document}
\bibliographystyle{apsrev4-1}

\title{Anomalous interfacial temperature profile induced by phonon localization}

\author{Yue Liu}
\author{Dahai He}\email{dhe@xmu.edu.cn}

\affiliation{Department of Physics and Institute of Theoretical Physics and Astrophysics, Xiamen University, Xiamen 361005, Fujian, China }

\date{\today}
\begin{abstract}
  Through the integration of the power spectral density, we obtain temperature profiles of both multi-segment harmonic and anharmonic systems, showing the presence of an anomalous negative temperature gradient inside the interfacial segment. Via investigating patterns of the power spectral density, we found that the counterintuitive phenomenon comes from the presence of interfacial localized phonon modes. Two out-band localized modes of the harmonic model, which make no contributions to local temperature due to the absence of phonon interactions, result in the concave temperature profile and over-cooling effect. For the anharmonic model, thanks to the phonon-phonon interactions, the localized modes are excited and make considerable contributions to interfacial temperature, which is clearly shown by examining the temperature accumulation function. When anharmonicity is considerably large, the negative temperature gradient is absent since the localized phonon modes are fully mixed. The presence of localized modes are evidently demonstrated by the inverse participation ratio and normal mode analysis for the isolated harmonic model.
\end{abstract}

\maketitle
\section{INTRODUCTION}\label{INTRODUCTION}
   In general, there will be a temperature discontinuity at an interface between two different materials when heat is transported across the interface. This effect, first observed at the interface between a metal and liquid helium~\cite{KAPITZA41}, was characterized by interfacial thermal resistance (Kapitza resistance). Two phenomenological models, namely, acoustic mismatch model~\cite{Little59} and diffuse mismatch model~\cite{Swartz87,Swartz89}, have been developed to estimate the interfacial thermal resistance. However, these models lack accuracy that can sometimes be deviated by orders of magnitude since they neglect the built-in atomic structures of interface and corresponding interfacial phonon states~\cite{Ramiere16}.

   When feature dimensions in micro/nanoelectronics continue to shrink, interfacial thermal resistance becomes one of the critical factors determining the performance of nanodevices. Particularly, as characteristic lengths of materials approach the energy carriers¡¯ mean free paths, the transport processes are no longer dominated by scattering inside the bulk materials, but are dominated instead by scattering at heterojunction interfaces. Recently, there have been extensive studies on interfacial thermal conduction through systems of atomic structures via analytical calculations~\cite{Saltonstall13,Polanco14,Wang07,Zhang11,Su15} and non-equilibrium molecular dynamics (NEMD) simulations~\cite{Xu14,Saaskilahti14,Saaskilahti16}. Meanwhile, experimental studies have shown that interfacial structures and bonding can have a marked effect on thermal transport~\cite{Pernot10,Sadeghi13,Duda13,Cahill-Rev14,Volz-Rev16}.  Nevertheless, previous studies have mainly focused on the interfacial thermal resistance but the temperature jump alone from a microscopic aspect has been rarely explored. When an interface with built-in atomic structures is concerned, quantitative understanding of the temperature discontinuity remains an unsolved issue.

   Recently, an interesting phenomenon called ``negative temperature jump'' has been observed in one-dimensional inhomogeneous harmonic and $\phi^{4}$ chains using the Langevin equations and Green's function (LEGF) method and NEMD simulations~\cite{Cao15}. The negative temperature jump nontrivially indicates the violation of Fourier's law in the interfacial part of systems. It has been shown that the existence of the anomalous interfacial temperature profile is not resulted from the system integrability or the sharp discontinuity of particles' interactions at interface. For the latter a multi-segment model has been proposed, enabling proper resolution of the molecular structure of the interface with a gradual variation of the molecular interactions. Despite the presence of the negative temperature jump has been verified from several aspects, a satisfactory explanation to elucidate its  microscopic origin has yet been absent.

   In this paper, we gain access to the underlying physical mechanism of the anomalous interfacial temperature profile in multi-segment chains of oscillators. In term of Parseval's theorem~\cite{Gradshteyn00}, we establish the relation between local temperature and the power spectral density (PSD). Temperature profiles can then be evaluated though the integration of the PSD and anomalous negative temperature gradient is recovered in the interfacial segment. By investigating the patten of PSD~\cite{Fu15}, we observe the presence of localized phonon modes of systems, which results in the occurrence of negative temperature gradient. The significant role played by anharmonicity is then investigated. The presence of phonon localization are evidently demonstrated by inverse participation ratio (IPR)~\cite{Chaudhuri10,Monthus10} and normal mode analysis for the harmonic system isolated from heat baths.

   The paper is organized as follows. In Sec.~\ref{sec2}, we introduce the model and methods used in this study. Particularly, we build the relation between PSD and temperature. In Sec.~\ref{sec4}, we calculate temperature profiles via the integration of PSD and explain the underlying physical mechanism of the negative temperature gradient through investigating the PSD pattern and normal modes of the system. Finally, we summarize our main conclusions in Sec.~\ref{sec4}.

\section{MODEL AND METHODS}\label{sec2}
    The present study is mainly to investigate temperature profiles of systems in a nonequilibrium steady state, for which we consider a one-dimensional chain of coupled atoms with nearest-neighbor interactions. The system isolated from heat baths is described by a Hamiltonian of the form
    \begin{eqnarray}\label{eq:1}
        \mathcal{H}=\sum\limits_{n = 0}^N \frac{p_{n}^{2}}{2m}+\frac{1}{2} k_{n}(x_{n+1}-x_{n})^{2}+U(x_{n}),
    \end{eqnarray}
    where $p_{n}$ and $x_{n}$ denote the instantaneous momentum and the displacement from the equilibrium position of $n$-th oscillator, respectively. $m$ is the mass of the particles. We use fixed boundary condition, i.e., $x_{0}=x_{N+1}=0$. $U(x)$ represents the $\phi^{4}$ on-site potential taking the form
    \begin{eqnarray}\label{eq:2}
        U(x)=\frac {1}{4}\lambda x^{4},
    \end{eqnarray}
    where $\lambda$ denotes the strength of the quartic potential. $k_{n}$ is the coupling strength of the $(n+1)$-th particle to the $n$-th particle. As mentioned above, what we mainly concerned is the temperature profile inside an interface of  particular microscopic structure between two molecular junctions. For this purpose, we consider a three-segment system of smoothly varied coupling strengths inside the interfacial segment~\cite{Cao15}. The distribution of coupling strengths of the chain is given by the following way: $k_{n}=1$ for $0\leq n< n_{1}$;  $k_{n}=1+\exp{({-\frac{(n-N/2)^2}{l}})}$ for $n_{1}\leq n\leq n_{2}$; and $k_{n}=1$ for $n_{2}< n\leq N$. The interface is modeled by the small intermediate segment with particles indexed from $n_1$ to $n_2$. The setup of smoothly-varying coupling strengths is to avoid the sharp discontinuity of temperature jump at interface, for which one can get more details of the ``temperature jump'' and the corresponding interfacial thermal resistance.

    The isolated system reduces to a multi-segment harmonic chain (MHC) when $\lambda=0$, for which one can use the normal mode analysis. In this case, normal modes of the harmonic lattice can be obtained by the following equation
    \begin{eqnarray}\label{eq:3}
        \mathbf{K}\mathbf{e_{p}}=\omega_{p}^{2}\mathbf{M}\mathbf{e_{p}},
    \end{eqnarray}
    where $\mathbf{M}$ is the mass-matrix of the system, and $\omega_{p}$ and $\mathbf{e_{p}}$ represent the eigenfrequency and eigenvector of $p$-th mode. Since we use the fixed boundary condition, the force-matrix of the system $\mathbf{K}$ (Hessian matrix) can be written as
    \begin{eqnarray}\label{eq:4}
        \mathbf{K}=
        \begin{bmatrix}
        k_{0}+k_{1} & -k_{1}     &   0        &\cdots     & 0           \\
        -k_{1}      &k_{1}+k_{2} &-k_{2}      &\cdots     & 0           \\
          0         &-k_{2}      &k_{2}+k_{3} &\ddots     &\vdots       \\
        \vdots      &\vdots      &\ddots      &\ddots     &-k_{N-1}     \\
          0         & 0          &\cdots      &-k_{N-1}   &k_{N-1}+k_{N}\\
        \end{bmatrix}.
    \end{eqnarray}

    For a given mode, the degree of localization can be characterized by the IPR~\cite{Chaudhuri10}
    \begin{eqnarray}\label{eq:5}
        P^{-1}= \frac {\sum\limits_{n} a_{n}^4}{(\sum\limits_{n} a_{n}^2)^2},
    \end{eqnarray}
    where $a_{n}$ is $n$-th component of the eigenvector of a given mode. For a completely delocalized state, $P^{-1}$ takes the value $\frac {1}{N}$. On the other hand, $P^{-1}$ takes the value $1$ for a fully localized state.

    When there exists anharmonicity ($\lambda\neq0$), the normal mode analysis fails. One can generally apply direct NEMD simulations to study properties of thermal transport through the multi-segment anharmonic chain (MAC). In this case, the particles at two ends (i.e. $n=1$ and $n=N$) are connected to the Langevin heat baths~\cite{Lepri-Rev03,Dhar-Rev08} at temperatures $T_{L}$ and $T_{R}$, respectively. The equations of motion are given by
    \begin{eqnarray}\label{eq:6}
        m\ddot{x}_{n}=-\frac{\partial \mathcal{H}}{\partial x_{n}}-\gamma_{n}\dot{x}_{n}+\eta_{n},
    \end{eqnarray}
    where $\gamma_{n}=\gamma(\delta_{n,1}+\delta_{n,N})$ and $\eta_{n}=\eta_{L}\delta_{n,1}+\eta_{R}\delta_{n,N}$. The noise terms $\eta_{L,R}$ denote a Gaussian white noise with zero mean and variance of $2\gamma k_{B}T_{L,R}$ according to the fluctuation-dissipation theorem, where $\gamma$ is the friction coefficient and $k_{B}$ is the Boltzmann constant. The local temperature is defined by $T_{n}=\langle\frac {p_{n}^2}{m}\rangle$~\cite{Lepri-Rev03,Dhar-Rev08}, for which the notion $\langle\cdots\rangle$ denotes the long-time average. In order to compute the temperature profile, the equations of motion (Eq.(\ref{eq:6})) are integrated using a stochastic Verlet algorithm. In our following discussions, the friction coefficient $\gamma$, the Boltzmann's constant $k_{B}$ and mass $m$ are set as units, i.e., $\gamma=1,k_{B}=1,m=1$.

    Since NEMD simulations cannot identify the spectral contributions of phonon modes to local temperature, we hereby present a numerical approach to quantify the temperature profile based on its spectral property, which is described by the PSD with respect to the momentum. The PSD can be conveniently calculated through the Fourier transform:
    \begin{eqnarray}\label{eq:7}
        S_{n}(\omega)=\left|\lim\limits_{\tau \to +\infty}\frac{1}{\tau}\int_{0}^\tau p_{n}(t)e^{-i\omega t} dt\right|^2.
    \end{eqnarray}
    According to Parseval's theorem~\cite{Gradshteyn00}, we have
        \begin{eqnarray}\label{eq:8}
        \frac {1}{\tau}\int_{0}^\tau \left|p_{n}^2(t)\right|dt=\frac {1}{\tau}\int_{0}^\infty \left|\int_{0}^\tau p_{n}(t)e^{-i\omega t} dt\right|^2 d\omega.
    \end{eqnarray}
    By taking $\tau \to +\infty$ and noting that $p_{n}(t)$ is real and positive, one get
    \begin{eqnarray}\label{eq:9}
        \lim\limits_{\tau \to +\infty} \frac {1}{\tau}\int_{0}^\tau p_{n}^2(t)dt=\int_{0}^\infty S_{n}(\omega)d\omega.
    \end{eqnarray}
    The left side of Eq.~\eqref{eq:9} is the time average of the kinetic energy $\langle p_{n}^2(t)\rangle$. According to the definition of temperature and noting that $m=1$, we have
    \begin{eqnarray}\label{eq:10}
        T_{n}=\int_{0}^\infty S_{n}(\omega)d\omega.
    \end{eqnarray}
    Eq.~\eqref{eq:10} can be regarded as the spectral decomposition of temperature, which indicates that the local temperatures are equivalent to the integration of the PSD at local sites, respectively. The local PSD $S_{n}(\omega)$ defined in Eq.~\eqref{eq:7} can be  evaluated numerically using the fast Fourier transformation. In our calculations, the time series' length of the instantaneous momentum $p_{n}(t)$ is $2^{19}$ and the sample interval $\Delta=1000h$ with the time step $h=0.001$ for the integration. PSD is averaged over 1000 realizations to reduce numerical fluctuations.

    The advantage of this approach is that one can not only obtain the temperature profile basing on PSD but also identify the spectral contributions of phonon modes to local temperature, particularly for anharmonic systems. To clarify the spectral contributions, we further define a temperature accumulation function
    \begin{eqnarray}\label{eq:11}
    T_{n}^{\textrm{accum}}(\omega)=\int_{0}^\omega S_{n}(\omega')d\omega',
    \end{eqnarray}
    which identifies the local temperature due to the accumulation of phonon modes with frequency less than $\omega$.
  \begin{figure}[h]
    \centering
    \includegraphics[width=1\linewidth]{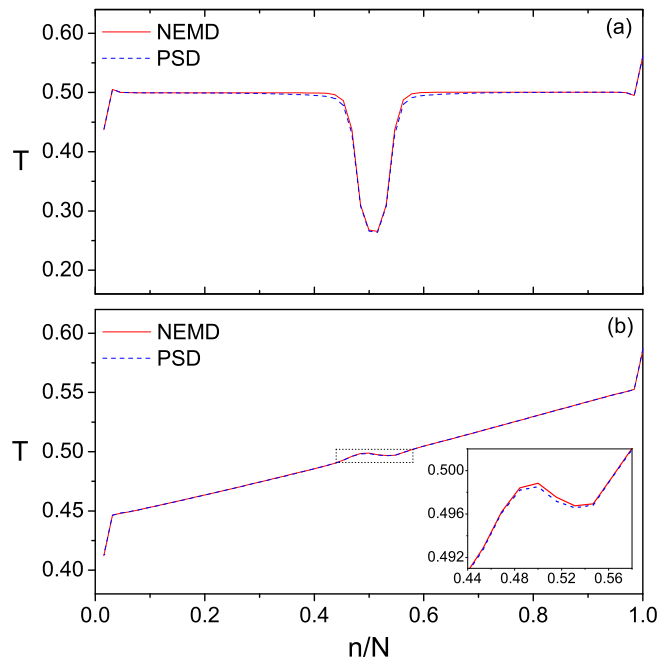}
    \caption{\label{Fig1}The temperature profiles evaluated by both the PSD approach (Eq.~\eqref{eq:10}) and NEMD simulations for (a) the multi-segment harmonic chain (MHC) and (b) the multi-segment anharmonic chain (MAC). For the two cases, $\lambda$ is equivalent to $0$ and $0.5$, respectively. Inset in (b) gives the enlarged plot for the interfacial temperature profile enclosed by the dotted rectangle. One can see a nice consistence between the PSD approach and NEMD simulations. The distribution of spring constants for the system is given by: $k_{n}=1$ for $0\leq n< 7N/16$; $k_{n}=1+\exp{(-8(n-N/2)^2/25)}$ for $7N/16\leq n\leq 9N/16$; and $k_{n}=1$ for $0< n\leq N$. Here $T_{L}=0.4$, $T_{R}=0.6$, and $N=64$.}
  \end{figure}

\section{RESULTS AND DISCUSSIONS}\label{sec3}
    According to Eq.~\eqref{eq:10}, temperature profiles obtained in terms of the integration of PSD are shown in Fig.~\ref{Fig1} for both MHC and MAC. As one can see from Fig.~\ref{Fig1}(a), the so-called ``negative temperature jump''~\cite{Cao15}, namely, an anomalous negative temperature gradient against the direction of heat flow, is recovered here shown in the concave interfacial temperature profile of MHC. Interestingly, the interfacial segment is anomalously ``over-cooled'' for the harmonic model, i.e., temperatures inside the central part are much lower than  $T_{L}$. As a comparison, temperature profile for the $\phi^{4}$ chain is illustrated in Fig.~\ref{Fig1}(b), where negative temperature jump still exists in the interfacial segment when $\lambda=0.5$. However, the concave degree of the temperature profile is smaller than that for the harmonic chain. The validity of the PSD approach is examined by the nice consistence with the results obtained directly by NEMD simulations.
  \begin{figure}[h]
    \centering
    \includegraphics[width=1\linewidth]{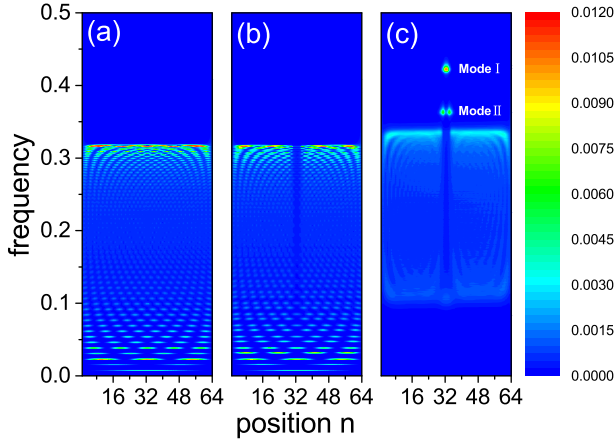}
    \caption{\label{Fig2}Patterns of the PSD for (a) the homogeneous harmonic chain (HHC), (b) MHC, and (c) MAC, respectively. The horizontal and vertical coordinates represent the index of particles along chain and frequency, respectively. The color represents the magnitude of PSD. For the case of HHC, the coupling strength is given by $k_{n}=1$. For the case of MHC and MAC, parameters are given the same as that for Fig.~\ref{Fig1}.}
    \end{figure}

    As to understand the existence of the anomalous temperature gradient, we analyze the PSD through identifying the contributions of phonon modes to temperature profiles. Fig.~\ref{Fig2} shows the patten of PSD for both MHC and MAC with respect to frequencies and lattice sites. As a comparison, we also evaluate PSD for the homogeneous harmonic chain (HHC), where the coupling strengths $k_{n}=1$ for all sites. From Fig.~\ref{Fig2} (a) and (b), one can see that the PSD for the interfacial segment of MHC is smaller than that of HHC, particularly in the middle- and high-frequency region. Thus temperature in the interfacial segment of MHC is less than that for HHC according to Eq.~\eqref{eq:10}. As we will see below, there are two phonon modes out of the allowed-frequency range (see Fig.~\ref{Fig6} below) are fully localized and forbidden to transport due to the absence of phonon interactions. This means that the localized modes make no contribution to local temperature, which results in the over-cooling effect. However, the two phonon modes (modes I and II) are delocalized (partly extended) in the presence of anharmonicity ($\lambda=0.5$), which is illustrated in Fig.~\ref{Fig2}(c). In this case, two out-band modes are excited in the middle of the interfacial segment and make considerable contributions to temperature thanks to the phonon-phonon interactions. Note that the phonon band (the allowed-frequency range) of the $\phi^4$ model moves up with a cut-off frequency (see Fig.~\ref{Fig2}(c)), which is caused by the presence of the on-site potential and can be approximately evaluated by the self-consistent phonon theory~\cite{He08,He09}.

\begin{figure}[h]
    \centering
    \includegraphics[width=1\linewidth]{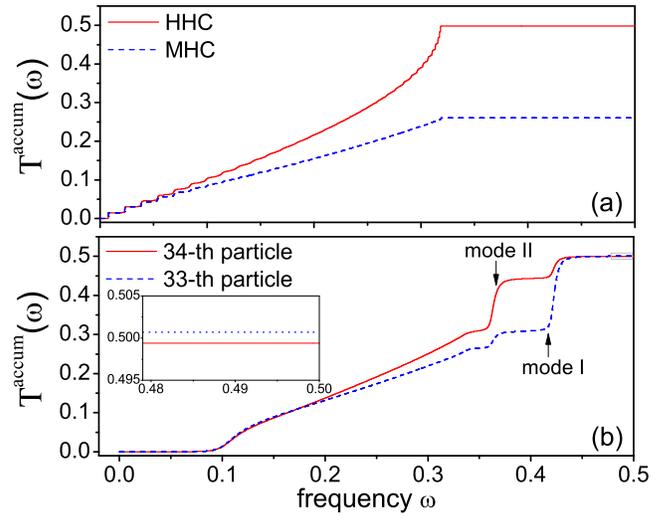}
    \caption{\label{Fig3}Comparing temperature accumulation function $T^{\textrm{accum}}(\omega)$ for (a) the 32-th particle of HHC and MHC and (b) the 33-th and 34-th particles of MAC, respectively. Inset: Enlarged plot for the region enclosed in the dotted rectangular, showing that $T_{34}<T_{33}$. The parameters are the same as that for Fig.~\ref{Fig1}.}
    \end{figure}

    The contribution of delocalized modes can be clearly shown by the temperature accumulation function $T^{\textrm{accum}}(\omega)$ in terms of Eq.~\eqref{eq:11}. In Fig.~\ref{Fig3}(a), we plot the temperature accumulation function for the 33-th particle of MHC and HHC, respectively. As we can see, the contribution of high-frequency modes are significantly suppressed in the presence of localized modes. Thus the temperature of the 33-th particle in MHC is much less than that in HHC. The same analysis can be applied to other particles and one can see that the concave temperature profile of MHC comes from the presence of interfacial localized modes that are not involved in thermal transport. Similarly, we plot $T^{\textrm{accum}}(\omega)$ for the 33-th and 34-th particles of MAC in Fig.~\ref{Fig3}(b). As mentioned above, the presence of anharmonicity leads to the delocalization of the two localized modes and makes them play a part in thermal transport through phonon-phonon interactions. The sharp increment of $T_{33}^{\textrm{accum}}(\omega)$ and $T_{34}^{\textrm{accum}}(\omega)$ comes from the contribution of mode I and mode II, respectively. Since the low-frequency mode is easier to be delocalized and thus more extended in the chain than the high-frequency one (see Fig.~\ref{Fig2}(c)),  Mode I has more concentrated energy and correspondingly larger PSD than mode II. Thus the temperature of the 33-th particle is higher than that of the 34-th particle, which indicates the presence of the negative temperature gradient.

\begin{figure}[h]
    \centering
    \includegraphics[width=1\linewidth]{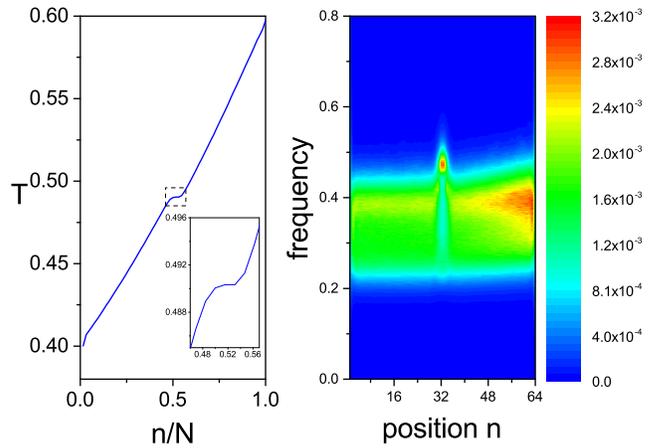}
    \caption{\label{Fig4} (a) Temperature profile for the multi-segment $\phi^4$ chain with strong anharmonicity $\lambda=10$. Inset is the enlarged plot for the interfacial temperature profile enclosed in the dashed rectangle, showing the absence of the negative temperature gradient. (b) The corresponding pattern of PSD. The color represents the magnitude of PSD. The horizontal and vertical coordinates represent the index of particles along chain and frequency, respectively. Other parameters except $\lambda$ are given the same as that for Fig.~\ref{Fig2}(c).}
    \end{figure}

    When anharmonicity increases, the phonon-phonon interactions increase, resulting in the enhancing mixing of the localized modes with in-band modes. For the case of strong anharmonicity ($\lambda=10$), the delocalized modes mix with each other intensively and have more extensive distribution of the PSD from the center of the system, as is illustrated in Fig.~\ref{Fig4}(b). The difference of their contributions to local temperature comes to diminish. Thus the negative temperature gradient inside the interfacial segment is absent (see Fig.~\ref{Fig4}(a)).

\begin{figure}[h]
    \centering
    \includegraphics[width=1\linewidth]{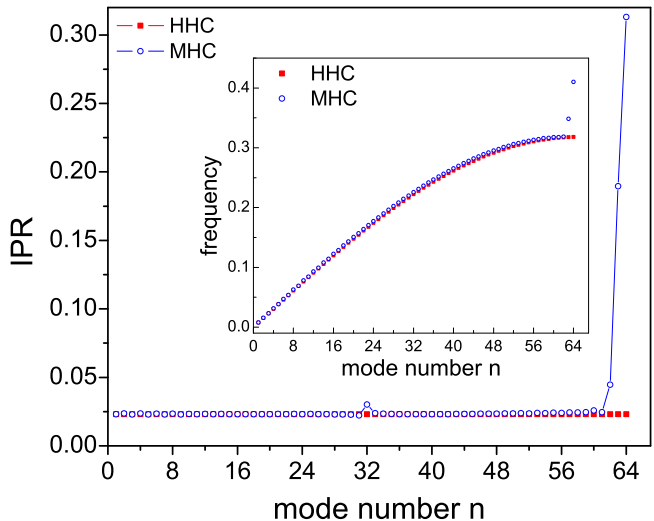}
    \caption{\label{Fig5}Inverse participation ratio for MHC and HHC. Inset: Dispersion relations for MHC and HHC, which shows that the last two normal modes have frequencies outside the bands of HHC. This corresponds to the large magnitude of the inver participation ratio for the last two normal modes. }
    \end{figure}

    Localized states occur in a solid when the translational invariance is lost. For gaining insights into the interfacial localized modes, we apply the normal mode analysis to MHC through the diagonalization of Eq.(\ref{eq:3}) and then obtain the eigenvalues and eigenvectors. Inset of Fig.~\ref{Fig5} shows that there are indeed two normal-mode frequencies are displaced outside the phonon band of HHC. The particular frequencies that experience this displacement are associated with localized modes whose space dependent factor decreases rapidly with distance from the middle parts.

    In order to measure the degree of localization, we calculate the IPR for MHC using Eq.~\eqref{eq:5}, as shown in Fig.~\ref{Fig5}. As a comparison, we also plot IPR for HHC. In the case of HHC, the IPR of every normal mode has the same value, as is expected. Note that the value is slightly larger than $\frac {1}{N}$, which comes from the finite-size effect. For MHC, the IPR of low-frequency normal modes are closely equal to the IPR of HHC since the low-frequency modes with wave-lengths longer than the size of system are extended. The IPRs of the last two normal modes, which have out-band frequencies, are much larger than others, corresponding to highly localized states.

    We plot some typical eigenvectors of normal modes for both MHC and HHC, as illustrated in Fig.~\ref{Fig6}. Note that we plot square of the wave amplitudes $a_{n}^2$ in stead of $a_{n}$ here in order to avoid showing the negative $a_{n}$. The eigenvector of the first normal mode of MHC are identical with that of HHC since the long wave-length mode of MHC are fully extended and the interface is ``transparent'' for the mode (see Fig.~\ref{Fig6}(a)). For the 56-th mode, a modulation of the amplitude at the interfacial sites can be seen. In other words, the mode is no longer a ``perfect''plane wave since the translational symmetry is broken (see Fig.~\ref{Fig6}(b)).  Fig.~\ref{Fig6}(c) and (d) show that the 63-th and 64-th normal modes are completely localized inside the interfacial segment, whose space-dependent amplitude decreases rapidly with distance from the center of the interfacial segment. This coincides with the above analysis on IPR. The two normal modes of MHC play significant roles in understanding the anomalous temperature gradient. Due to the absence of phonon-phonon interactions, the out-band phonon modes are completely localized inside the interfacial segment and can not be excited in the process of thermal transport. Moreover, defected wave amplitudes in interfacial segment (see Fig.~\ref{Fig6}(b)) corresponds to small PSD in middle-frequency region, leading to the concave temperature profile and the negative temperature gradient. The two modes still ``survive'' in the presence of small anharmonicity (corresponding to mode I and mode II, respectively) although they are partly delocalized. Although the modes are outside the band, they are excited in the process of thermal transport via the interactions of phonons. The components of their eigenvector are large for the interfacial segment, which results in large PSD and negative temperature gradient.
    \begin{figure}[h]
    \centering
    \includegraphics[width=1\linewidth]{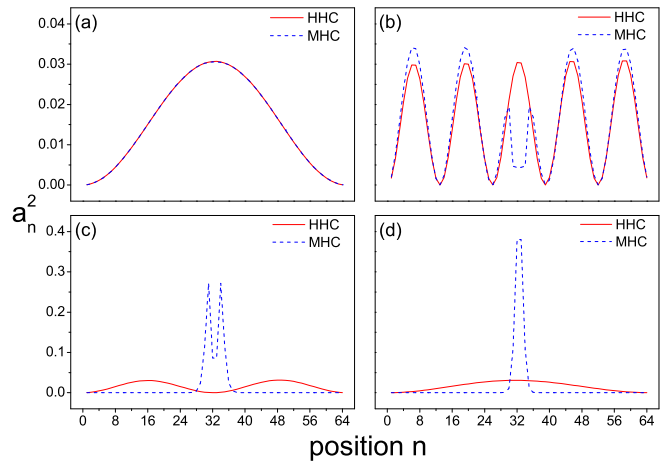}
    \caption{\label{Fig6}Square of wave amplitudes of (a) the first, (b) the 56-th, (c) the 63-th, and (d) the 64-th normal modes for both MHC and HHC. The localization of wave amplitudes for the 63-th and 64-th normal modes is evident.}
    \end{figure}
\section{SUMMARY}\label{sec4}
    In summary, we studied temperature profiles of the multi-segment harmonic and anharmonic systems. In terms of Parseval's theorem, one is capable of establishing the relation between temperature and PSD and then obtaining the spectral decomposition of temperature. Through the PSD approach, we recovered the counterintuitive temperature profile in the interfacial segment, where the local temperature gradient is anomalously against the direction of heat flow. By inspecting the PSD patterns and identifying the contributions of phonon modes, one can understand that the concave temperature profile and the over-cooling effect in MHC come from the presence of out-band localized modes, which are forbidden to transport and make no contribution to local temperature. However, for anharomonic systems, those modes are (partly) delocalized and make contributions to local temperature thanks to phonon-phonon interactions, which can be seen by investigating the temperature accumulation function. With the increase of anharmonicity, the localized modes come to mix with the in-band modes, leading the vanishing of the negative temperature gradient. The calculation of IPR and normal mode analysis evidently show the localization of two out-band phonon modes, which is consistent with the PSD analysis.

    Our further simulations shows that similar anomalous temperature profile can be observed in other multi-segment anharmonic systems, e.g., Fermi-Pasta-Ulam-$\beta$ lattice. Note that the spectral decomposition of temperature can also be obtained by the LEGF method~\cite{Cao15}. However, it cannot be applied to anharmonic systems in comparison with the PSD approach. Furthermore, the PSD analysis can be applied to disordered anharmonic systems~\cite{Dhar-Rev08,Wang15,Wang16} and possibly help to understand the interplay between anharmonicity and disorder. Nanotechnology nowadays has the ability to manipulate the interfacial structure and bonding, which allows for a gradual changing of interfacial properties~\cite{Cahill-Rev14,Volz-Rev16}. For example, recent experiments have demonstrated that inserting a thin Ti layer at Au/Si interfaces~\cite{Duda13} or bridging the metal-carbon-nanotube interface with covalently bonded organic molecules~\cite{Kaur14} can dramatically enhance the interfacial thermal conductance. By elucidating the origin of the anomalous temperature profile, our study is expected to enlighten further understanding the Kapitza resistance from the microscopic viewpoint and possible designing nanoscale devices for heat manipulation.

    \begin{acknowledgments}
     We acknowledge the helpful discussions with Bei-Lok Hu, Hong Zhao, Jiao Wang, and Yong Zhang. Financial support from NSFC of China (Grant Nos. 11675133 and 11335006), NSF of Fujian Province (No. 2016J01036), and President Grant of Xiamen University (No. 20720160127) is acknowledged.
    \end{acknowledgments}
%
\end{document}